\documentclass[%
superscriptaddress,
amsmath,amssymb,
prl,
aps,
10pt,
twocolumn
]{revtex4-1}

\usepackage[english]{babel}
\usepackage[utf8]{inputenc}
\usepackage{graphicx}
\usepackage[caption=false]{subfig}
\usepackage{float}
\usepackage{dcolumn}
\usepackage{blindtext}
\usepackage{upgreek}
\usepackage[squaren]{SIunits}
\usepackage{xcolor}

\newcommand{\ket}[1]{\left|#1\right\rangle}
\newcommand{\kcirc}[1]{\ket{#1 \mathrm{C}}}
\newcommand{\kell}[1]{\ket{#1 \mathrm{E}}}

\begin{document}

\title{Laser Trapping of Circular Rydberg Atoms}

\author{R.~G.~Corti\~nas}
\altaffiliation{These authors contributed equally to this work.}
\author{M.~Favier}
\altaffiliation{These authors contributed equally to this work.}
\author{B.~Ravon}
\author{P.~M\'ehaignerie}
\author{Y.~Machu}
\author{J.~M.~Raimond}
\author{C.~Sayrin}
\email[Corresponding author: ]{clement.sayrin@lkb.ens.fr}
\author{M.~Brune} 
 
\affiliation{Laboratoire Kastler Brossel, Coll\`ege de France, CNRS, ENS-Universit\'e PSL, Sorbonne Universit\'e, 11 place Marcelin Berthelot, F-75231 Paris, France}

\date{\today}

    \begin{abstract}
Rydberg atoms are remarkable tools for quantum simulation and computation. They are the focus of an intense experimental activity mainly based on low-angular-momentum Rydberg states. Unfortunately, atomic motion and levels lifetime limit the experimental time-scale to about $100\,\micro\second$. Here, we demonstrate laser trapping of long-lived circular Rydberg states for up to $10\,\milli\second$. Our method is very general and opens many opportunities for quantum simulation. The $10\,\milli\second$ trapping time corresponds to thousands of interaction cycles in a circular-state-based quantum simulator. It is also promising for quantum metrology and quantum information with Rydberg atoms.   
    \end{abstract}

\maketitle    
    
Rydberg atoms are blessed with remarkable properties. Their long lifetimes, large sensitivities to external fields and huge dipole-dipole interactions have led to milestone developments in quantum simulation~\cite{Weimer2010, Labuhn2016, Bernien2017, Zeiher2017}, quantum information~\cite{Saffman2010}, quantum optics~\cite{Lukin2003, Mohapatra2007, Peyronel2012, Baur2014, Gorniaczyk2014, Distante2016, Busche2017, Ripka2018, Li2019}, quantum sensing~\cite{Sedlacek2012a, Facon2016, Cox2018} and molecular physics~\cite{Bendkowsky2009, Hollerith2019b}. Most of these experiments use laser-accessible low-angular-momentum states, with a lifetime in the hundred microsecond range.

Circular Rydberg Atoms (CRAs), with maximal angular momentum ($\ell = |m| = n-1$, $n$ being the principal quantum number) have a much longer intrinsic lifetime, $30\,\milli\second$ for $n=50$. This makes them ideal tools for cavity quantum electrodynamics~\cite{Haroche2013} or quantum-enabled sensing~\cite{Dietsche2019, Facon2016}. 
Moreover, CRAs are utterly promising for quantum simulation~\cite{Nguyen2018}. Tens of laser-trapped circular atoms, protected from spontaneous emission~\cite{Hulet1985}, interacting through a fully tailorable dipole-dipole interaction can simulate the long-time evolution of a complex interacting spin system.

Laser trapping of CRAs~\cite{Dutta2000} is instrumental in the realization of this quantum simulator, but would also benefit spatially-resolved quantum-enabled sensing. For low-angular-momentum Rydberg states, early experiments have demonstrated laser-~\cite{Anderson2011, Li2013} and electric-field~\cite{Seiler2011, Zhelyazkova2019} trapping. Recently, optical tweezers for low-$\ell$ Rydberg states have been reported~\cite{Barredo2019}, with interesting perspectives for quantum simulation. However, the short lifetime of these levels and their high photoionization rate~\cite{Saffman2005} limit the trapping to a few hundred~$\micro\second$.

Here, we demonstrate laser trapping of circular Rydberg atoms. We excite laser-cooled atoms to the $n=52$ circular Rydberg level inside a hollow Laguerre-Gauss (LG) laser beam. These atoms are repelled by the laser-induced ponderomotive potential~\cite{Dutta2000} and transversally confined in the light tube for times up to $10\,\milli\second$. The trapping frequency is measured to be $1.37\,\kilo\hertz$. We check that the lifetime and the coherence of the trapped atoms are not sensibly affected by the light, as expected for levels insensitive to photoionization~\cite{Nguyen2018} and for a nearly $n$-independent trapping potential~\cite{Dutta2000}.

\begin{figure*}[!t]
    \includegraphics[width = \textwidth]{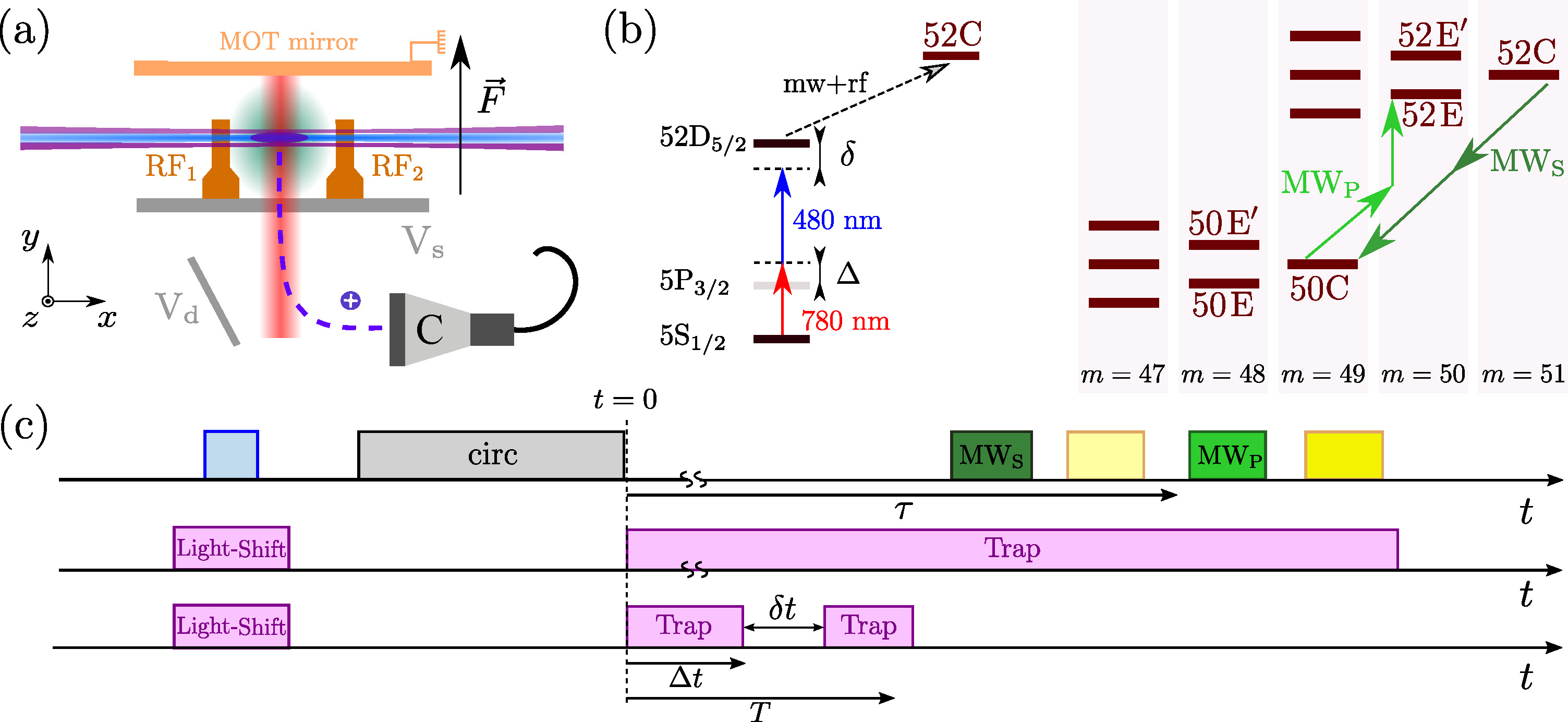}
    \caption{\textbf{(a) Experimental setup} with axes definition. The blue and red Rydberg excitation lasers cross in the cold atom cloud (green), about $4\,\milli\meter$ away from the surface of the MOT mirror. The electrode V$_\mathrm{s}$ applies the electric field $F$ and the ionization field. The Rb$^+$ ions (violet dashed line) are guided towards the channeltron C by the deflection electrode V$_\mathrm{d}$. Four additional electrodes (only two, RF$_1$ and RF$_2$, are shown for clarity) apply the circular state preparation rf field and a static electric field gradient $\partial_z F$. The LG beam (purple) traps the Rydberg atom cloud (dark blue)
    \textbf{(b) Simplified level scheme.} Left: two-photon laser excitation (solid arrows) and circular state preparation (dotted arrow). Right: Partial diagram of the Stark levels in the $n=50$ and $n=52$ manifolds sorted by $m$ values. The arrows represent the microwave transitions MW$_\mathrm{S}$ (level selection) and MW$_\mathrm{P}$ (probe).
    \textbf{(c) Timing of the experiment} (not to scale). Top frame: main events. From left to right: laser excitation (light blue), circular state preparation (grey), selection pulse MW$_\mathrm{S}$ (dark green), partial ionization (light yellow), probe pulse MW$_\mathrm{P}$ (dark green), field ionization (dark yellow). Middle frame: LG beam intensity versus time for the trapping experiment. Bottom frame: LG intensity for the trap oscillation-frequency measurement.}
    \label{pic:Setup_scheme}	
\end{figure*}

The experimental setup is shown on Fig.~\ref{pic:Setup_scheme}(a). Ru\-bi\-di\-um-87 atoms are laser-cooled and trapped in a 3D mirror-magneto-optical trap (MOT) formed in front of a Rubidium-coated mirror~\cite{Hermann-Avigliano2014,Teixeira2015} enclosed in a $4\,\kelvin$ $^4$He cryostat that shields the Rydberg atoms from room-temperature black-body radiation~\cite{Suppl}. The MOT is loaded from a 2D-MOT atomic beam propagating along the $z$-direction [axes definition in Fig.~\ref{pic:Setup_scheme}(a)]. A molasses stage further cools the atoms down to $\simeq 10-20\,\micro\kelvin$ in their ground state $\ket{5\mathrm{S}_{1/2}, F=2}$. 

We excite the atoms from $\ket{5\mathrm{S}_{1/2}, F=2}$ into $\ket{52\mathrm{D}_{5/2}, m_J=5/2}$ by a two-photon laser excitation [Fig.~\ref{pic:Setup_scheme}(b)]. It makes use of a $780\,\nano\meter$-wavelength red laser beam ($100\,\micro\meter$ diameter) perpendicular to the MOT mirror and of a $480\,\nano\meter$-wavelength blue laser beam ($22\,\micro\meter$ diameter) aligned with the $x$-axis. The two crossed beams define a cigar-shaped Rydberg cloud elongated along the $x$-direction. The red laser is blue-detuned from the intermediate $\ket{5\mathrm{P}_{3/2}, F'=3, m_{F'}=3}$ level by $\Delta = 560\,\mega\hertz$, while the frequency of the blue laser can be scanned around the two-photon resonance condition. Both lasers are shaped in a $2\,\micro\second$-long pulse.

The excitation is performed in an $F=0.8\,\volt\per\centi\meter$ electric field along the $y$-axis, defined by the grounded MOT mirror and electrode V$_\mathrm{s}$ [Fig.~\ref{pic:Setup_scheme}(a)]. After the end of the laser pulses, we adiabatically transfer the atoms into the $n=52$ circular state ($m=+51$, noted $\kcirc{52}$), using microwave (mw) and radiofrequency (rf) transitions~\cite{Hulet1983, Suppl, Cortinas2019}. 
We measure the populations of individual Rydberg levels via state-selective field ionization. We apply with V$_\mathrm{s}$ a $200\,\micro\second$-long electric field ramp, which successively ionizes the Rydberg levels.  The measurement distinguishes low-$m$ levels from high-$m$ levels with the same $n$ and resolves levels with different $n$s~\cite{Cortinas2019}. The resulting ions are counted in separate time windows. The circular state purity, deduced from microwave spectroscopy, is $>90\%$~\cite{Suppl}. The remaining population is distributed over a few high-$m$ elliptical states ($n = 52, m \lesssim n-2$).

We trap the CRAs in a ponderomotive trap acting on the almost-free Rydberg electron~\cite{Dutta2000, Suppl}. 
Rydberg states are low-field seekers and are trapped in the local intensity minimum on the axis of a Laguerre-Gauss LG$_{01}$ beam. We use a $1064\,\nano\meter$-wavelength fiber laser~\cite{ALS} and tailor it with a spatial light modulator (SLM)~\cite{SLM} into a LG beam~\cite{Suppl}. The laser beam is sent onto the atoms along the $x$-axis [Fig. \ref{pic:Setup_scheme} (a)].

An intensity profile of the incoming trapping beam recorded outside the cryostat only gives a qualitative insight into the intensity distribution at the location of the atomic cloud. Three successive thick glass windows on the cryostat vacuum tank and thermal shields appreciably distort the incoming beams. We probe \emph{in situ} the trap beam intensity profile by measuring the light shifts it induces on the $\ket{5\mathrm{S}_{1/2}} \rightarrow \ket{52\mathrm{D}_{5/2}}$ ground-Rydberg transition~\cite{Suppl}. Getting spectra for several positions of the trapping beam w.r.t.~the blue beam, we reconstruct the LG beam intensity distribution. We use this information to optimize the LG-beam shape, correcting with the SLM the aberrations introduced by the cryostat windows. Finally, the LG beam is found to be slightly elliptical with waists of $35\,\micro\meter$ and $41\,\micro\meter$ along the $z$- and $y$ axes, respectively. The maximum light shift of the Rydberg excitation line is $3.8\,\mega\hertz$. The reconstructed peak laser intensity is thus $I=6.6\cdot10^4\,\watt\per\centi\meter^2$ (total power $4.0\,\watt$), corresponding to a $80\,\micro\kelvin$ trap depth, much larger than the initial atomic temperature.

The timing of the experiment demonstrating the circular atoms trapping is sketched in Fig.~\ref{pic:Setup_scheme}(c). We perform the Rydberg excitation in presence of the trapping beam, switched on for $15\ \micro\second$ around the $2\ \micro\second$-long excitation pulse. We precisely center the LG beam on the axis of the blue laser by minimizing the observed light shift. Due to the finite size of the blue beam, the minimum average light shift is about 2 MHz. Setting $\delta = 0.5\,\mega\hertz$, where $\delta$ is the two-photon detuning from resonance, the excitation selects, within the diameter of the blue laser, only atoms close the center of the LG beam. We estimate the transverse size of the initial Rydberg cloud to be $\sim 10\,\micro\meter$~\cite{Suppl}. We switch off the trapping beam and perform in about $30\,\micro\second$ the transfer to the circular state, completed at time $t=0$. 

We then compare the spatial expansion of the Rydberg cloud with and without the LG trapping beam, optionally switched on at $t=0$. After a time delay $\tau$, varying between $0.8$ and $9.8\,\milli\second$, we probe the Rydberg cloud spatial extension by mw spectroscopy. We monitor the broadening of a probe transition to a neighboring manifold, which experiences a linear Stark effect in an electric field gradient. It maps the Rydberg atoms positions onto their resonance frequencies. 

For the larger $\tau$ values, blackbody-induced transfer of population from the initial $\kcirc{52}$ state to adjacent circular states (trapped as well in the LG beam) is significant. These transfers create a spurious background for the probe transition spectroscopy. In order to get rid of this background, we selectively transfer at time $\tau-600\ \micro\second$ all $\kcirc{52}$ atoms (irrespective of their position) into $\kcirc{50}$ by a `hard' $0.8\,\micro\second$-long microwave selection $\pi$-pulse MW$_\mathrm{S}$. After an additional $150\,\micro\second$ delay, we apply, during $160\,\micro\second$, a partial ionization field ramp. It ionizes all Rydberg atoms with $n>50$ and does not affect the population of $\kcirc{50}$. We apply, at $\tau-200\,\micro\second$, a strong electric field gradient along the $z$-axis, $\partial_z F = 0.56\,\volt\per\centi\meter^2$, using the RF electrodes [Fig.~\ref{pic:Setup_scheme}(a)]. The gradients in the other directions are $\partial_x F = 0.18\,\volt\per\centi\meter^2$ and $\partial_y F = 0.10\,\volt\per\centi\meter^2$. We let the field relax to its steady-state ($F=1.46\,\volt\per\centi\meter$) at $t=\tau$.

Finally, we apply at $t=\tau$ a $1.8\ \micro\second$-long probe mw pulse MW$_\mathrm{P}$ on the $\kcirc{50}\rightarrow\kell{52}$ transition. The level $\kell{52}$ is the low-lying $\ket{n, m=n-2}$ elliptical level, experiencing a first-order Stark shift of 99.8~MHz/(V/cm) (see level scheme in Fig.~\ref{pic:Setup_scheme}). A gaussian fit to the spectrum of this probe transition determines its Full Width at Half Maximum (FWHM), $\sigma_\mathrm{P}$. 

\begin{figure}[!t]
  \includegraphics[width=\columnwidth]{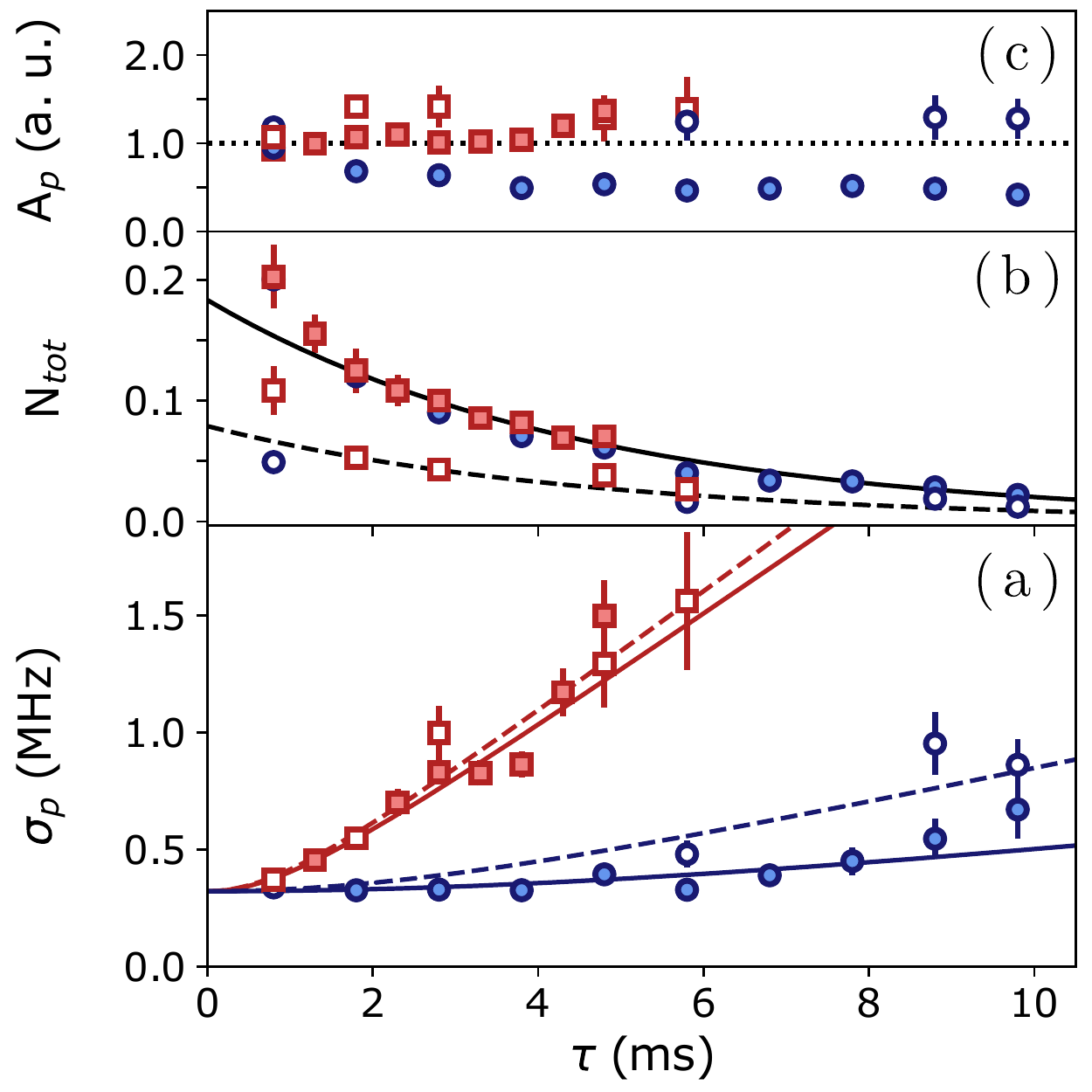} 
    \caption{{\bf Laser trapping.} (a) FWHM $\sigma_\mathrm{P}$ of the probe transition as a function of the delay time $\tau$ for untrapped (red rectangles) and trapped (blue circles) atoms. In all panels, full symbols correspond to an average atom number $\bar{N}_0$. Open symbols correspond to the same measurements with about $\bar{N}_0/2$ atoms. The lines result from a model of the Rydberg cloud expansion for untrapped (red) and trapped (blue) atoms, with $\bar{N}=\bar{N}_0$ (solid) or $\bar{N}\approx\bar{N}_0/2$ (dotted). (b) Total population of $\kcirc{50}$ at $t=\tau$, before the MW$_\mathrm{P}$ probe pulse for trapped (blue circles) and untrapped (red rectangles) atoms. The lines correspond to an exponential fit with a single decay time of ($4.6 \pm 0.3\,\milli\second$). The number of atoms at $t=0$, uncorrected from detection efficiency, is $\bar{N} = \bar{N}_0 = 0.18 \pm 0.01$ for solid symbols and $\bar{N} = 0.08 \pm 0.01 \approx \bar{N}_0/2$ for open symbols. (c) Integrated area $A_\mathrm{P}$ of the MW$_\mathrm{P}$ probe spectrum for trapped (blue) and untrapped (red) atoms. Error bars correspond to 1-$\sigma$ standard error deviation.}
    \label{pic:Trapping}%
\end{figure}

Figure~\ref{pic:Trapping}(a) presents $\sigma_\mathrm{P}$ as a function of $\tau$ with (blue circles) and without (red squares) the trapping beam, and for two average Rydberg atom numbers in the sample at $t=0$, $\bar N = \bar{N}_0$ (solid symbols) and $\approx\bar{N}_0/2$ (open symbols). Without trapping light, the Rydberg cloud thermally expands because of its finite temperature. When turning on the trapping laser, the linewidth remains basically constant. We only observe a slow broadening at long times resulting from the motion of the atoms along the unconfined $x$-axis in the electric field gradient $\partial_x F$. 

The lines present a fit to the data with the predictions of a 3D (red) or 1D (blue) expansion model~\citep{Suppl}. For the untrapped case, it is in excellent agreement with the observed data with a temperature of $T\approx 13.5\,\micro\kelvin$, independently of $\bar{N}$. For the low-$\bar{N}$ trapped-atoms case (dotted blue line), we find $T = (14\pm 3)\,\micro\kelvin$. These temperatures are in good agreement with a direct time-of-flight measurement after the molasses stage. However, for the high-$\bar{N}$ trapped-atoms case (solid blue line), we find $T=(3\pm1)\,\micro\kelvin$. 
This lower value of $T$ may result from interactions between trapped CRAs, as explained below.
Note that we never observe, within experimental noise, a wide pedestal on the trapped-atoms narrow line, which would be the signature of an untrapped fraction~\cite{Suppl}. These data vividly demonstrate the main result of this paper: The atoms are trapped with a high efficiency and stay localized at the center of the trap for up to $10\,\milli\second$.

In Fig.~\ref{pic:Trapping} (b), we plot the total population in $\kcirc{50}$ at $t=\tau$, before MW$_\mathrm{P}$, for trapped (blue circles) and untrapped (red squares) atoms. In both cases, the population decay reflects the thermal transfers from the initial $\kcirc{52}$ state to neighboring circular states before the selection MW pulse (MW$_\mathrm{S}$). We observe no significant modification of the atomic lifetime when the trap is on. This shows that photoionization is quite negligible over this time scale, as expected~\cite{Nguyen2018}.

Furthermore, we check that the purity of the circular states is not appreciably affected by the trapping. To this end, we plot the integrated area $A_\mathrm{P}$ of the recorded probe microwave spectrum as a function of $\tau$, with and without trapping light. A reduced $A_\mathrm{P}$ value indicates a transfer of population to other levels in the same manifold, which are not addressed by MW$_\mathrm{P}$. The area $A_\mathrm{P}$ thus measures the purity of the circular levels, independently of their position in the electric field gradient. The results are shown in Fig.~\ref{pic:Trapping}(c) for $\bar N = \bar{N}_0$ and $\approx\bar{N}_0/2$. For the lowest atom number, $A_\mathrm{P}$ is nearly constant for untrapped and trapped atoms, revealing that the trapping light does not affect the circular Rydberg level purity. 

For the highest atom number, $A_\mathrm{P}$ is also constant in the untrapped case but decays rapidly (within $4\,\milli\second$) to about half of its initial value for trapped atoms. 
This effect could be explained by the interactions between the trapped atoms. The 1D trap enhances the probability that two atoms come close enough to collide. These collisions may redistribute population in the Stark manifold. This would happen at a higher rate for faster atoms, explaining the rapid initial decay of $A_\mathrm{P}$ for the high $\bar{N}$ value, after which we expect that mainly slow atoms remain in $\kcirc{52}$. It is consistent with the smaller effective temperature observed as compared to the lower $\bar{N}$ case. The low expansion temperature in Fig.~\ref{pic:Trapping}(a) (solid blue circles) can thus be viewed as the harbinger of interatomic interactions of CRAs in the laser trap.

As a final check, we measure the CRAs trapping frequency~\cite{Nogrette2014}. The timing of the experiment is shown in Fig.~\ref{pic:Setup_scheme}(c). We move the trapping beam by $12\,\micro\meter$ along the $z$-axis w.r.t.\ the blue laser. As before, the LG beam is on for $15\,\micro\second$ during Rydberg excitation. Setting the two-photon laser detuning to $\delta = 2.1\,\mega\hertz$, about half of the light shift at maximum LG intensity, we excite Rydberg atoms on the inner slope of the LG beam~\cite{Suppl}. After preparation of the circular states, we switch the trap on at time $t=0$. The atoms start an oscillation along the $z$-axis at the trap frequency.

In order to probe this oscillation, we turn off the LG beam for $\delta t = 300\,\micro\second$ after a variable waiting time $\Delta t$. Depending upon their velocity at $\Delta t$, the atoms can fly away from the trapping region. If it is not the case, we recapture them for a duration $T-\Delta t$, where $T = 2.6\,\milli\second$. We finally perform the detection sequence as above, setting MW$_\mathrm{P}$ at resonance with the $\kcirc{50}\to\kell{52}$ transition. The duration $T-\Delta t$ is chosen long enough so that the atoms flying away in the electric field gradient are impervious to MW$_\mathrm{P}$. It thus addresses only those atoms that remained trapped.

\begin{figure}[!t]
\includegraphics[width=8cm]{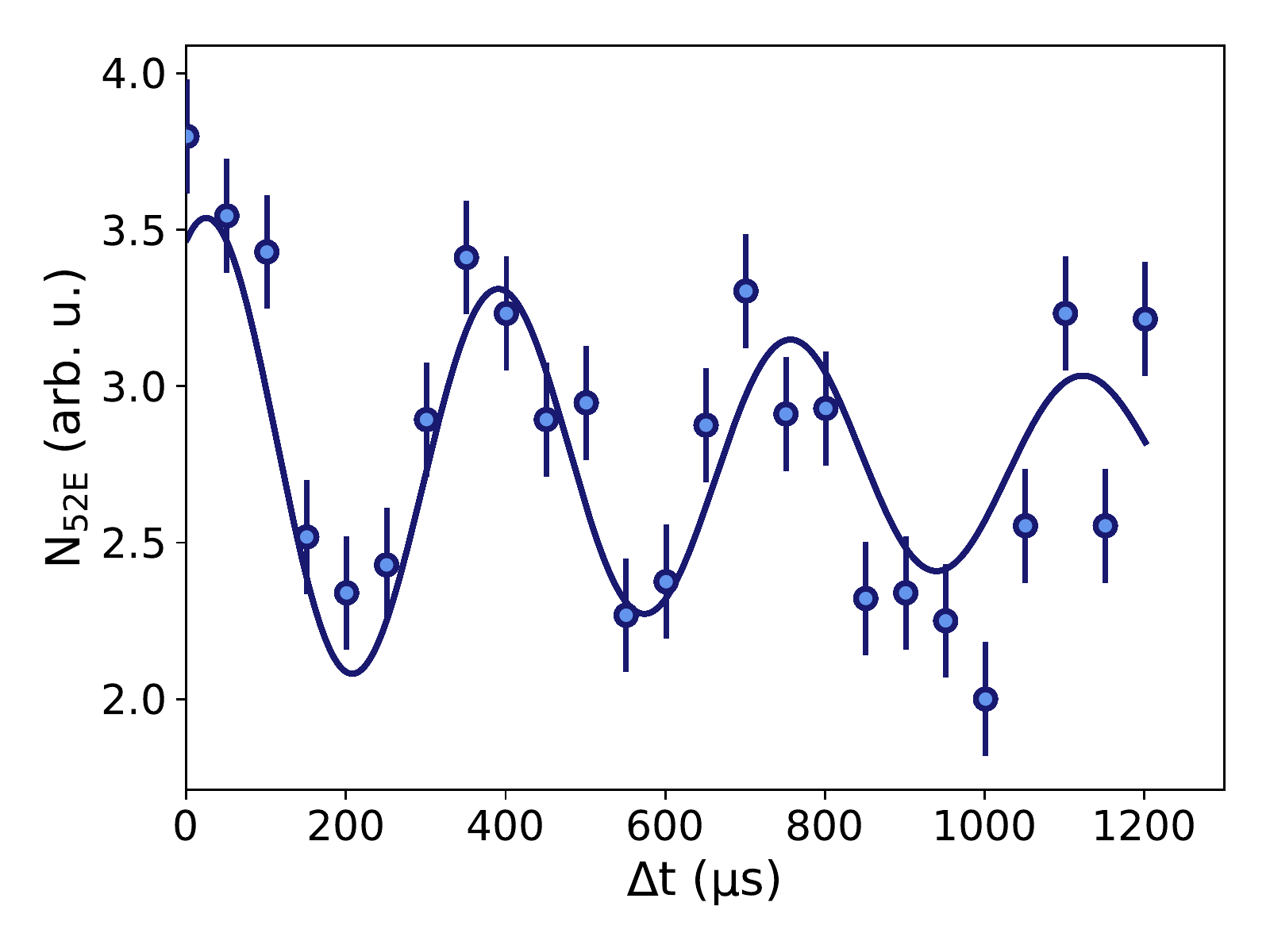}
    \caption{\textbf{Trap frequency measurement.} Blue circles with statistical error bars: number of atoms, $N_{{52\mathrm{E}}}$ in $\kell{52}$, as a function of the time $\Delta t$ at which the trap is switched off. Blue line: damped sinusoidal fit to the data. 
    }
    \label{pic:oscillation}%
\end{figure}

\begin{figure}[!t]
\includegraphics[width=8cm]{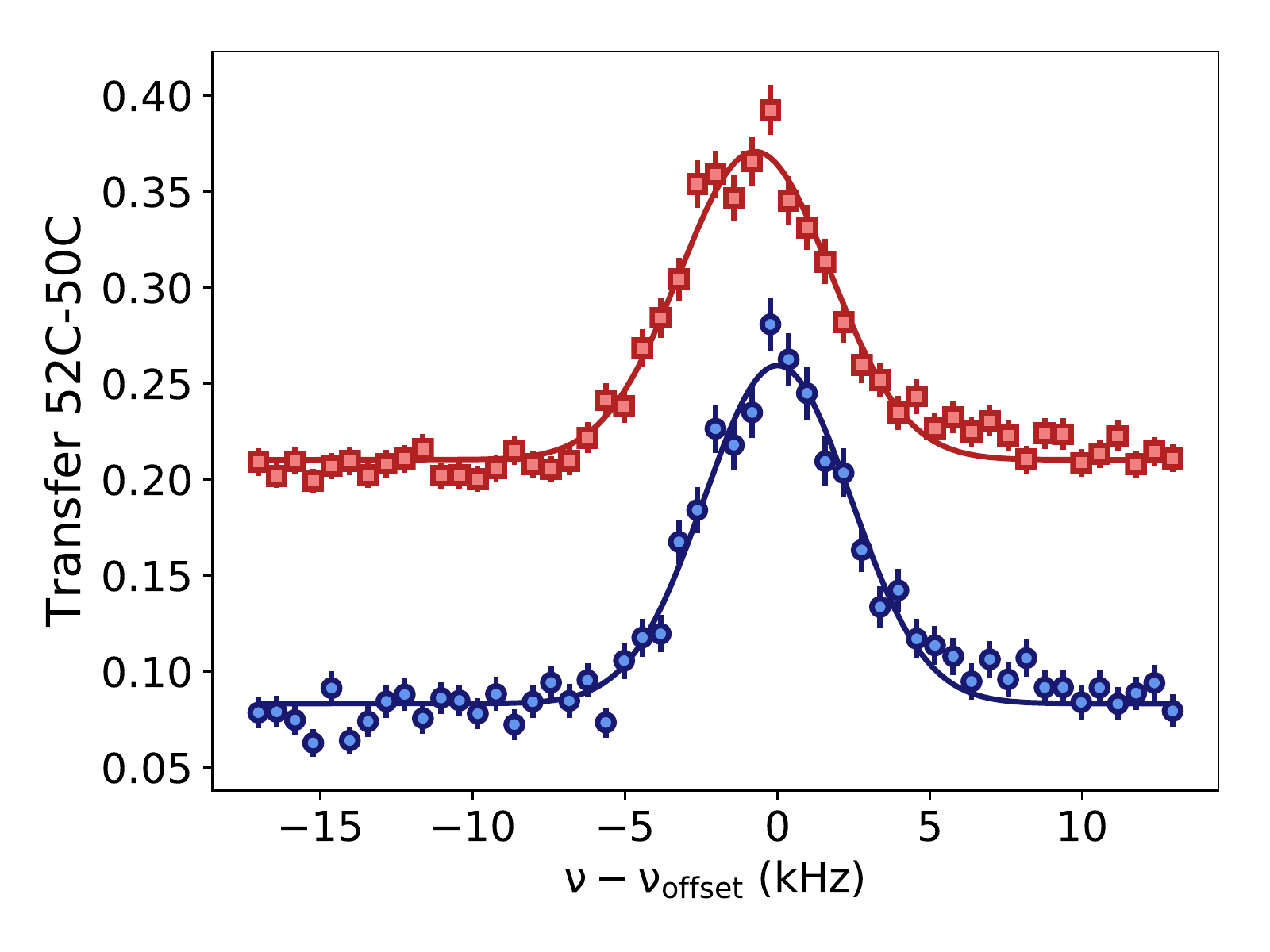}
    \caption{\textbf{Test of atomic coherence.} Microwave spectrum of the two-photon $\kcirc{50}\to\kcirc{52}$ transition for trapped (blue circles) and untrapped (red rectangles) atoms (statistical error bars) with $\nu_{\mathrm{offset}}=49.639071\ \giga\hertz$. The red points have been shifted upwards for clarity. }
    \label{pic:spectra}%
\end{figure}

In Fig.~\ref{pic:oscillation}, we plot (blue circles) the number $N_{{52\mathrm{E}}}$ of atoms in $\kell{52}$ as a function of $\Delta t$. We clearly observe damped oscillations. The recapture probability, and so $N_{{52\mathrm{E}}}$, is lowest when the atoms are released at the time they speed through the bottom of the trap, \emph{i.e.}\ when their kinetic energy is maximum. The population $N_{{52\mathrm{E}}}$ thus oscillates at twice the trapping frequency. From a fit of the signal to a damped sine (solid blue line), we extract a trap frequency of $(1.37\pm 0.05)\ \kilo\hertz$. It is in fair agreement with a Monte-Carlo simulation of the anharmonic atomic motion, which yields a frequency of $1.1\,\kilo\hertz$~\cite{Suppl}. 

Finally, we check the coherence of the trapped circular atoms. We record, without applied field gradients, the spectrum of the nearly electric-field insensitive $\kcirc{50}\to\kcirc{52}$ transition driven by a $215\ \micro\second$-long microwave pulse. Figure~\ref{pic:spectra} shows the spectra obtained for untrapped (red squares) and trapped (blue circles) atoms. We observe a small $(0.7\pm0.2)\,\kilo\hertz$ shift between the two situations, corresponding to a $1\,\milli\volt\per\centi\meter$ electric field drift over the one-hour data acquisition time, compatible with our experimental stability. The spectrum linewidth is probably determined by electric and magnetic field noises~\cite{Cortinas2019}. The FWHM of gaussian fits to the data (solid lines in Fig.~\ref{pic:spectra})  for untrapped and trapped atoms are respectively $(5.8\pm 0.2)\,\kilo\hertz$ and $(5.7\pm 0.3)\,\kilo\hertz$. We thus observe, at the $200\,\hertz$ level, no effect of the trapping on the transition coherence, as expected for a level-independent trapping potential.

We have demonstrated laser trapping of circular Rydberg atoms in two dimensions, for up to $10\,\milli\second$. This time scale is currently limited only by the atomic lifetime in a finite-temperature environment. We have characterized the trapping potential and verified that it affects neither the circular levels lifetimes nor their coherence properties. To the best of our knowledge, the laser-trapping time demonstrated here is unprecedented for Rydberg atoms. It amounts to 1000 exchange-interaction cycles between two circular Rydberg atoms, with $n\sim50$, held $5\,\micro\meter$ apart. This work is an important step towards quantum simulation over long time scales, relevant to problems such as quenches across quantum phase transitions and thermalization~\cite{Nguyen2018} and towards the development of quantum-enabled sensors based on CRAs.
 
\medskip
This work has been supported by the European Union FET-Flag project n\degree 817482 (PASQUANS), ERC Advanced grant n\degree\ 786919 (TRENSCRYBE) and QuantERA ERA-NET (ERYQSENS, ANR-18-QUAN-0009-04), by the Region Ile-de-France in the framework of DIM SIRTEQ and by the ANR (TRYAQS, ANR-16-CE30-0026).

\bibliography{bibliography} 
\end{document}


\maketitle

\section{Preparation of the circular Rydberg states}

\begin{figure*}[!bt]
	\centering
    \subfloat[]{{\includegraphics[width=0.6\textwidth]{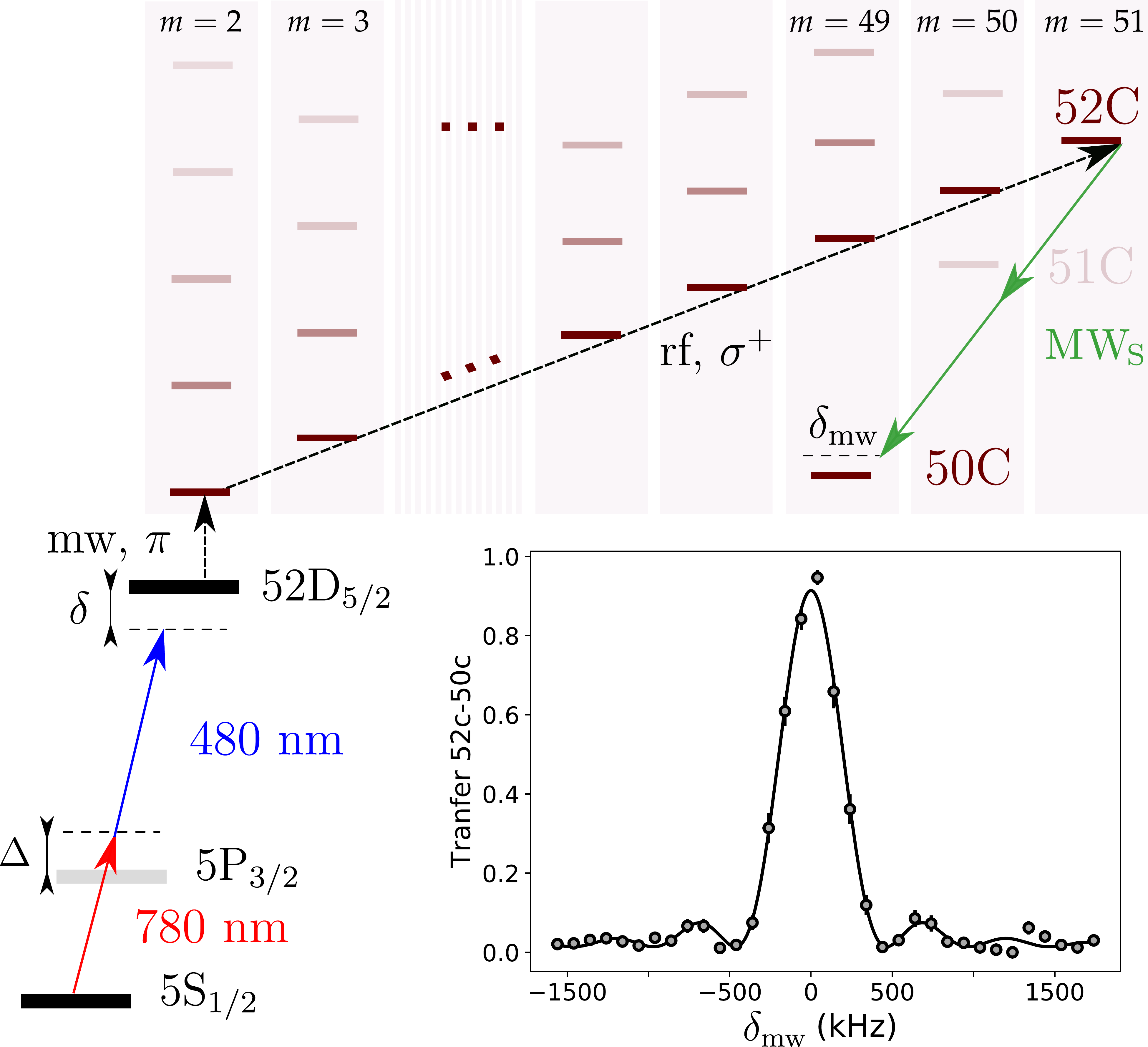} }}%
    \caption{Simplified level scheme for the circularization procedure. (Inset) Microwave spectroscopy of the two-photon $\kcirc{52}\to\kcirc{50}$ (MW$_\mathrm{s}$) transition. The solid line is a sinc$^2$ fit to the data. Error bars correspond to 1-$\sigma$ statistical errors.}
    \label{pic:Spectro_circ}%
\end{figure*}

We laser excite the Rubidium atoms from their ground-state $\ket{5\mathrm{S}_{1/2}, F=2}$ to the $\ket{52\mathrm{D}_{5/2}, m_J=5/2}$ level by two-photon laser excitation. The excitation is performed in an $F=0.8\,\volt\per\centi\meter$ electric field to lift the degeneracy between $m_J$ levels. A simplified level scheme is given in Fig.~\ref{pic:Spectro_circ}. We then adiabatically transfer the atoms to $\ket{52\mathrm{F}, m=2}$, applying a mw field at $57.7\,\giga\hertz$ during $4\,\micro\second$ while continuously varying the electric field across the resonance, which is hit at an electric field of $F=1.8\,\volt\per\centi\meter$.

Next, we switch the electric field to $F=2.4\,\volt\per\centi\meter$, turn on a $\sigma^+$-polarized radio-frequency field at $230\,\mega\hertz$ and ramp the electric field down to $F=2.1\,\volt\per\centi\meter$ in $2\,\micro\second$. An adiabatic rapid passage transfers the atoms from  $\ket{52\mathrm{F}, m=2}$ to $\kcirc{52}$. 
The rf field is produced by applying $230\,\mega\hertz$-frequency electric potentials on two out of the four RF electrodes (see Fig.~1, the two electrodes that are not shown are symmetric to RF$_1$ and RF$_2$ w.r.t.\ the $(x,y)$ plane). A precise control of the relative amplitudes and phases of these potentials is critical to minimize the $\sigma^-$-polarization component of the rf field.

\medskip 
We assess the purity of the prepared state $\kcirc{52}$ by performing mm-wave spectroscopy on the $\kcirc{52} \to \kcirc{50}$ transition with a $1.4\,\micro\second$-long $\pi$-pulse (Fig.~\ref{pic:Spectro_circ} inset). In the applied electric-field, this transition is well separated by $2.7\,\mega\hertz$ from the nearest transition starting from the $n=52$ elliptical states. The $91\%$-transfer from $\kcirc{52}$ to $\kcirc{50}$ at resonance is a lower-bound estimation of the purity of $\kcirc{52}$ (it assumes a unit-efficiency $\pi$-pulse). 

\section{\emph{In-situ} characterization of the trapping beam using light-shift measurements}

\subsection{Light shifts of the ground-Rydberg transition}
In order to reconstruct \emph{in-situ} the intensity profile of the LG trapping beam, we record the laser excitation spectrum of the $\ket{5\mathrm{S}_{1/2}, F=2} \to \ket{52\mathrm{D}_{5/2}, m_J=5/2}$ ground-Rydberg transition while shining the trapping light on the atoms. The trapping-laser light shifts both the ground- and the Rydberg-state energy levels by $h \beta_\mathrm{G} I$ and $h \beta_\mathrm{Ry} I$, respectively, where $I$ is the laser intensity and $h$ is the Planck constant. From the $[h\,(\beta_\mathrm{Ry}-\beta_\mathrm{G}) I]$-light shift of the transition, we infer $I$ at the position of the Rydberg atoms.

\medskip
The ponderomotive energy shift of the Rydberg energy levels reads
\begin{align}
	\mathcal{E} = h \beta_\mathrm{Ry} I = \frac{e^2}{2 m_e \epsilon_0 c \omega_L^2} I, 
\end{align}
where $e$ is the electron charge, $m_e$ its mass, $\epsilon_0$ the vacuum permittivity, $c$ the speed of light and $\omega_L$ the trapping laser angular frequency. 

\medskip

The scalar polarizability of the $\ket{5\mathrm{S}_{1/2}}$ ground state at a $1064\,\nano\meter$-wavelength
has been calculated by Safronova \emph{et al.}~\cite{Safronova2004} to be
\begin{align}
	\alpha_\mathrm{G} = 4\pi\epsilon_0 \times 693.5(9)\,a_0^3,
\end{align}
where $a_0$ is the atomic Bohr radius. The light shift of the ground state is then given by
\begin{align}
	\mathcal{E}_\mathrm{G} = h \beta_\mathrm{G} I = -\alpha_\mathrm{G}\, \frac{I}{2\epsilon_0 c}.
\end{align}
Numerically we find
\begin{align}
	\beta_\mathrm{Ry} &= 2.56 \cdot 10^{-5} \mega\hertz\per(\watt\per\centi\meter^2), \\
	\beta_\mathrm{G} &=  -3.25 \cdot 10^{-5}\, \mega\hertz\per(\watt\per\centi\meter^2).
\end{align}
We eventually get the total light-shift of the ground-Rydberg transition 
\begin{align}
	\Delta \nu &= \left(\beta_\mathrm{Ry}-\beta_\mathrm{G}\right)\, I = \beta_\mathrm{tot} I \approx 5.81 \cdot 10^{-5}\, \mega\hertz \, \frac{I}{\watt\per\centi\meter^2}.
\label{eq:LS_tot}
\end{align}

\medskip
The LG beam section is slightly elliptical due to residual optical aberrations, with principal axes oriented along the $y$- and $z$-axes. Its intensity at a position $(y,z)$ away from the beam axis, \emph{i.e.}\! the $x$-axis, reads
\begin{align}
	I(y,z)=P \frac{4}{\pi\,w_y\, w_z} \left(\frac{y^2}{w_y^2}+\frac{z^2}{w_z^2}\right) \exp\left[-2\left(\frac{y^2}{w_y^2}+\frac{z^2}{w_z^2}\right)\right],
\label{eq:LG_I}
\end{align}
where, $P$ is the power of the LG beam and $w_y$ and $w_z$ are the beam radii at the position $x$. The peak intensity is
\begin{align}
	I_\mathrm{max} = \frac{2 P}{\pi e \,w_y\, w_z}.
\label{eq:LG_I_max}
\end{align}

\subsection{Measurement of the LG beam intensity profile}
\label{sec:profile}
\begin{figure*}[!t]
	\centering
	\includegraphics[width=0.98\textwidth]{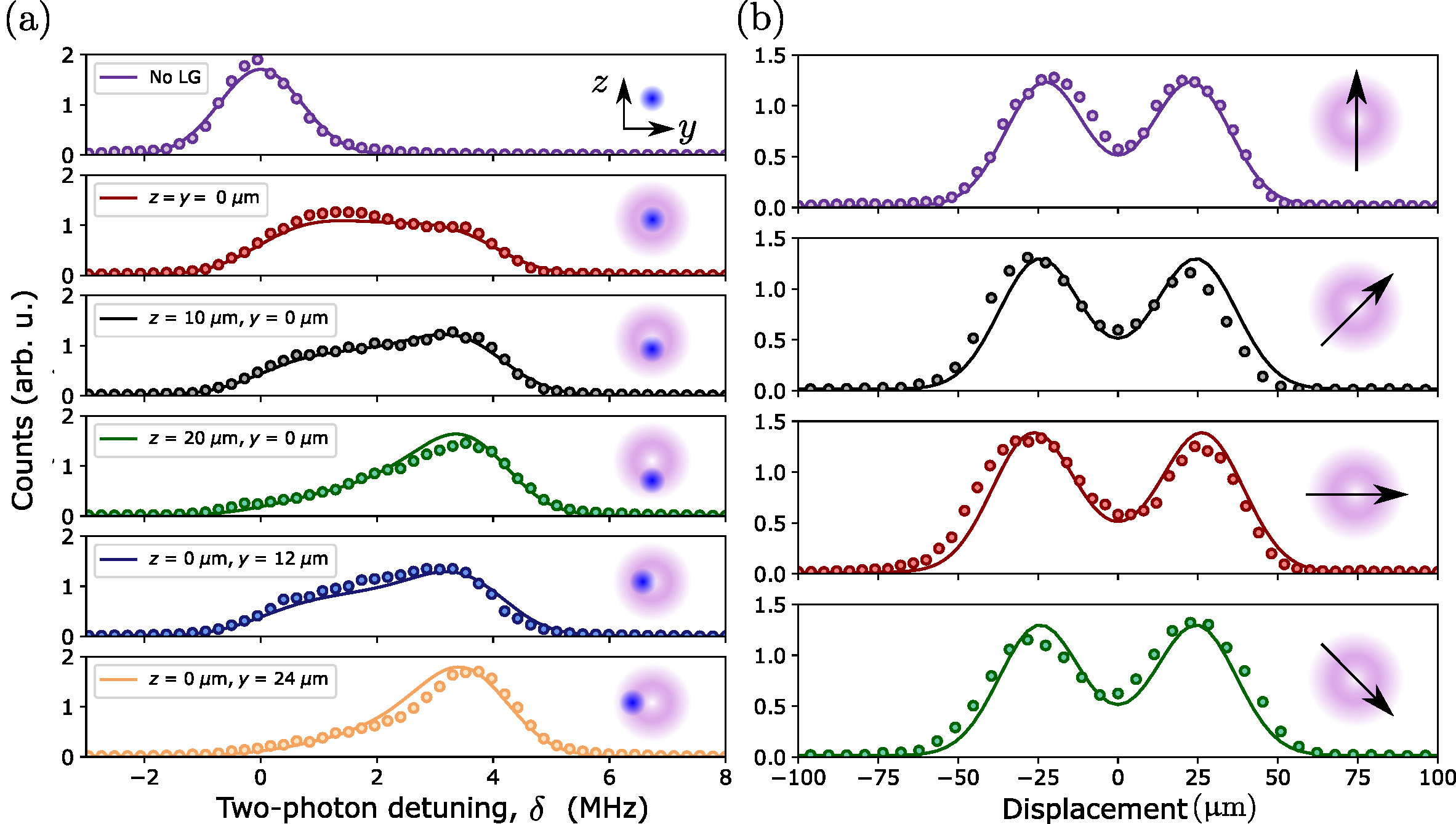}%
\caption{(a) Rydberg excitation spectra. First panel: The trapping laser beam is off during the laser excitation. Second to sixth panels: The trapping laser beam is on, its position with respect to the blue laser is different for each panel. The relative position of the two beams is given in the legend and schematically shown as an inset. (b) Number of excited Rydberg atoms as a function of the LG beam position w.r.t.\ the $x$-axis, at a fixed two-photon detuning of $\delta = 4.5\,\mega\hertz$. The direction of the LG beam displacement is given in the insets. In all panels, circles correspond to experimental results (error bars are smaller than the data points), while the solid lines are the result of a global fit to the data.}
    \label{pic:LS}%
\end{figure*}

We record the spectrum of the ground-Rydberg transition for several positions of the trapping beam w.r.t.~the blue beam (Fig.~\ref{pic:LS}(a), five bottom panels). We also plot the laser-excitation spectrum without trapping light (top panel, purple circles). Ground-state atoms are attracted towards the high-intensity rings of the LG beam. To avoid as far as possible this perturbing effect, the trapping laser is let on for $15\,\micro\second$ only during the Rydberg laser excitation, maintaining an homogeneous atomic density over the excitation region. When the finite-size blue laser is at the center of the LG trapping beam (red), the average light shift is minimal and is about $2\,\mega\hertz$. The other panels show the excitation spectrum with the LG beam shifted out of the $x$-axis, either along the $y$- or the $z$-axis. Moving the LG beam axis away from the $x$-axis shifts the excitation spectrum to higher frequencies.

\begin{figure*}[!t]
	\centering
	\includegraphics[width=0.45\textwidth]{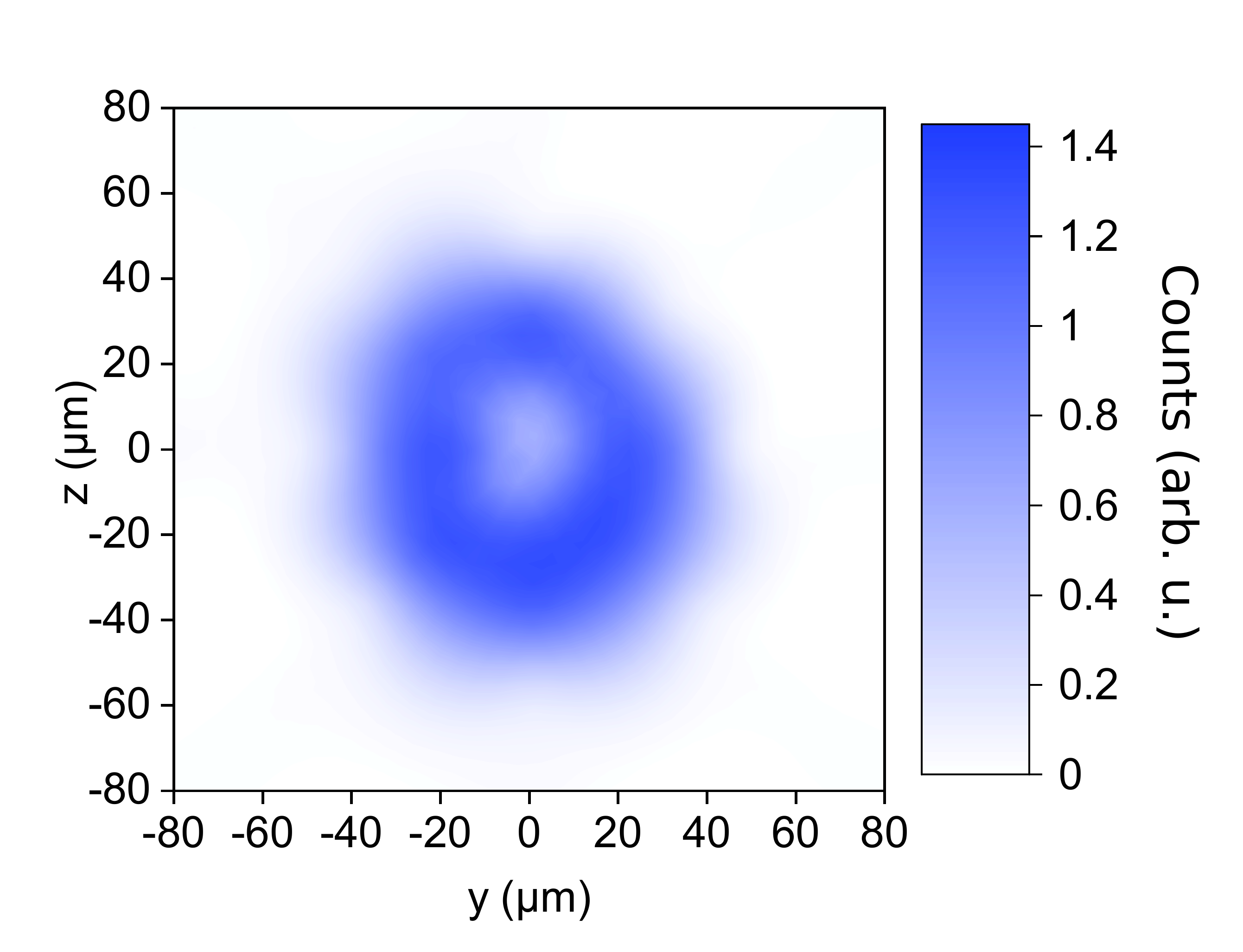} %
    \caption{In-situ qualitative reconstruction of the shape of the trapping beam using light-shift measurements. The 2D plot is an interpolation of the data of Fig.~\ref{pic:LS}(b)}
    \label{pic:LS2D}%
\end{figure*}

In Fig.~\ref{pic:LS}(b), we plot the number of atoms excited to $\ket{52\mathrm{D}}$ as a function of the LG beam position for a fixed two-photon detuning of $\delta = 4.5\,\mega\hertz$. It is chosen to be close to the maximum measured light-shift of the ground-Rydberg transition. In these experiments, the LG beam is scanned through the blue laser beam along four different directions: the $y$- and $z$ axes as well as the two diagonals tilted by $\pm 45\degree$ w.r.t.\ these axes. Figure~\ref{pic:LS2D} is a 2D plot of the measured data.

Were the linewidth $\gamma$ of the laser-excitation spectrum without trapping light negligible, figures~\ref{pic:LS}(b) and~\ref{pic:LS2D} would reveal the iso-intensity ($\beta_\mathrm{tot} I(y,z) = \delta$) lines of the LG beam. Instead, we observe their convolution with the gaussian-shaped spectrum in Fig.~\ref{pic:LS2D}(a), top panel. Atoms get excited to the Rydberg state only if the light-shift induced by the LG beam intensity at their position equals $\delta$, at a precision of $\gamma$. The ring structure of the LG beam is clearly visible. The minima at the center of these lines mark the point, at which the LG beam is aligned with the blue beam.

Asymmetries in the LG beam intensity profile, that could be induced by the thick optical ports of the cryostat, directly modify the reconstructed profile of Fig.~\ref{pic:LS2D}. We use these scans to image the aberrations and correct for them by adding correction phase masks on the SLM that tailors the trapping beam. Figures~\ref{pic:LS}(b) and~\ref{pic:LS2D} have been obtained after optimization.

\bigskip
We analyze the data in Fig.~\ref{pic:LS} by using a simple model. First, we fit the excitation spectrum without trapping light to a gaussian peak proportional to
\begin{align}
	\Gamma_\mathrm{B}(\delta) = \sqrt{\frac{2}{\pi \gamma^2}}\exp\left({-\frac{2\delta^2}{\gamma^2}}\right),
\end{align}
of FWHM $\gamma \sqrt{2 \ln(2)} = (1.67 \pm 0.05)\,\mega\hertz$. We then calculate the number $n(y,z)$ of excited atoms at a position $(y,z)$ via 
\begin{align}
	n(y,z) = n_0 \times \Gamma_\mathrm{B}\left[\delta - \beta_\mathrm{tot} I(y-y_\mathrm{LG},z-z_\mathrm{LG})\right] \times I_\mathrm{B}(y,z),
\end{align}
where $n_0$ is a normalization factor, $(y_\mathrm{LG}, z_\mathrm{LG})$ are the coordinates of the LG beam center and $I_\mathrm{B}(y,z)$ is the blue laser intensity. It reads
\begin{align}
	I_\mathrm{B}(y,z) = I_\mathrm{B}^{(0)} \exp\left[-\frac{2 (y^2+z^2)}{w_\mathrm{B}^2}\right],
\end{align}
where $w_\mathrm{B}$ is the waist of the blue laser and $I_\mathrm{B}^{(0)}$ its peak intensity. Note that the Rayleigh lengths of the blue and LG beams are large enough w.r.t.\ the Rydberg cloud size so that we can neglect the intensity variations along the $x$ axis.

The data in Fig.~\ref{pic:LS} is eventually fitted by 
\begin{align}
	N( \delta, y_\mathrm{LG},z_\mathrm{LG}) = \iint n(y,z) \,\mathrm{d}y \, \mathrm{d}z + N_0,
\end{align}
where $N_0$ accounts for a small offset in the experimental data. The free parameters of the model are $w_\mathrm{B}$, $w_y$, $w_z$, $I_\mathrm{max}$ and $N_0$. In addition, we allow for different values of $n_0$ for Fig.~\ref{pic:LS}(a) and Fig.~\ref{pic:LS}(b), taking into account the different mean number of excited atoms in the two experiments.

We obtain very good agreement with the experimental results for a maximum light-shift of $\Delta\nu = \beta_\mathrm{tot} I_\mathrm{max} = (3.81 \pm 0.06)\,\mega\hertz$ and a diameter of the blue laser of $w_\mathrm{B} = (21.9 \pm 0.8)\,\micro\meter$. The light-shift $\Delta\nu$ corresponds to $I_\mathrm{max} = 6.6 \cdot 10^4\, \watt\per\centi\meter^2$ and a trap depth of $80\,\micro\kelvin$. Given the LG beam waists of $w_y = (41\pm3)\,\micro\meter$ [$w_z = (35\pm1)\,\micro\meter$] along the $y$ [$z$] direction, we  predict an oscillation frequency at the bottom of the trap of $(1.12\pm0.08)\,\kilo\hertz$ [$(1.33\pm0.05)\,\kilo\hertz$]. We estimate a total power of $(4.0\pm0.3)\,\watt$ of trapping light, in good agreement with the measured $4.6\,\watt$ power of the beam that exits the cryostat.

Using these results, we eventually calculate the spatial distribution $n(y,z)$ of the Rydberg atoms in the experimental conditions of Figs.~2 and 3, \emph{i.e.}\ when the LG beam is aligned with the blue laser ($y_\mathrm{LG} = z_\mathrm{LG} = 0$) and $\delta = 0.5\,\mega\hertz$ and when the LG beam is displaced w.r.t.\ the blue laser ($y_\mathrm{LG} = 0$ and $z_\mathrm{LG} = 12\,\micro\meter$) and $\delta = 2.1\,\mega\hertz$. We find, in both cases, the diameter of the excited Rydberg cloud to be $\sim10\,\micro\meter$. 

\bigskip
It should be noted that the simple model that we use to reconstruct the trapping beam parameters has some limitations. Most importantly, it assumes an ideal shape for the LG beam, given in Eq.~\eqref{eq:LG_I}. In particular, it does not include possible fluctuations of the peak intensity along the ring of the trapping potential. Because of the finite size and spectral width of the blue laser, fast variations of the intensity are smoothed out and thus not detected by our reconstruction method. Nevertheless, this model  reproduces very well the data in Fig.~\ref{pic:LS}. It allows us to recover the main features of the trapping potential and, thus, to efficiently optimize \emph{in situ} the shape of the LG beam.

\section{Thermal expansion of the Rydberg cloud}
\begin{figure}[!ht]
\centering
	\includegraphics[width=0.5\textwidth]{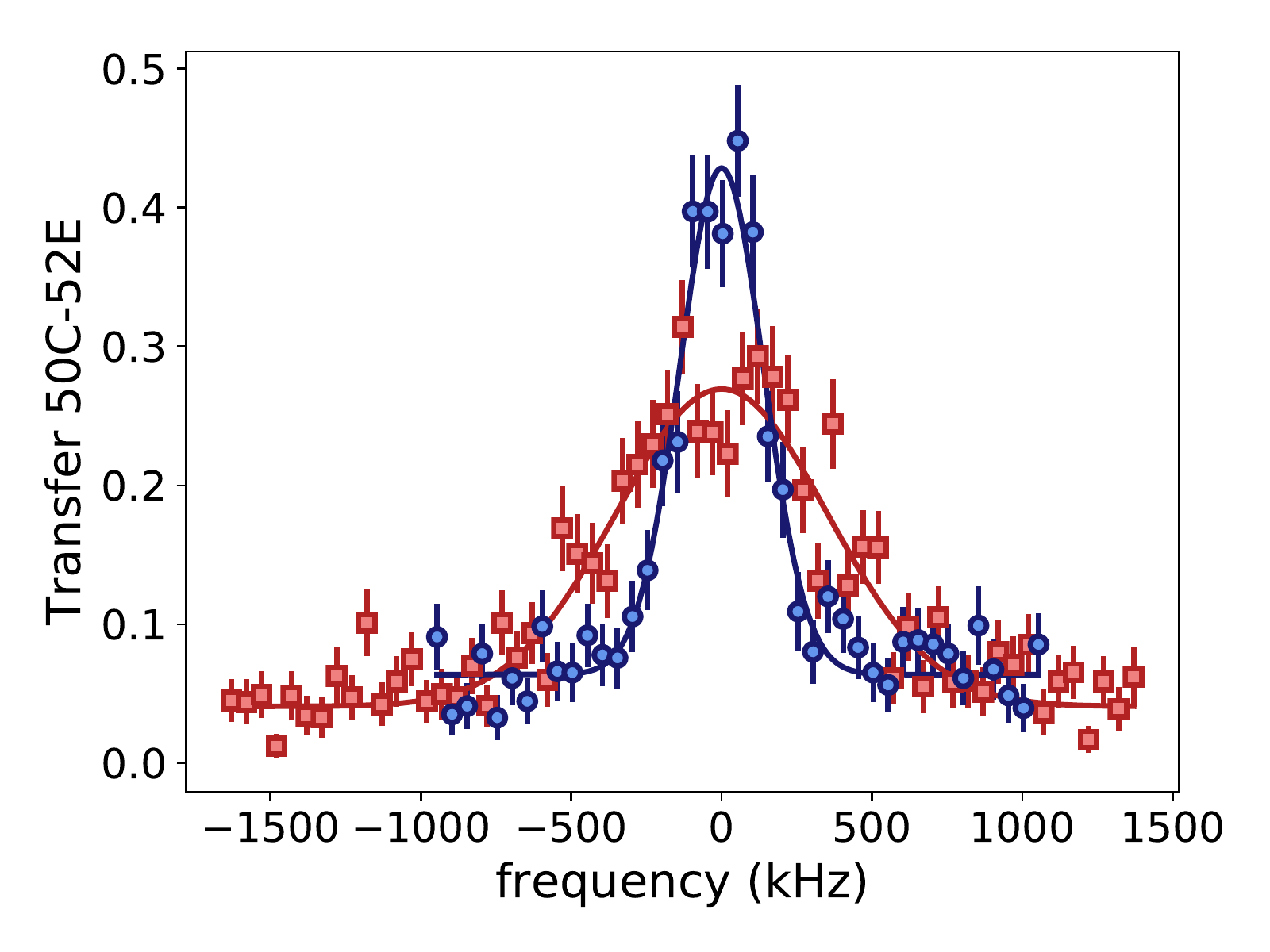}
\caption{Spectrum of the MW$_\mathrm{P}$ transition recorded with untrapped (red rectangles) and trapped (blue circles) atoms, after a free expansion of $\tau = 2.8\,\milli\second$. The solid lines are gaussian fit to the data.}
\label{pic:SpectroCE}
\end{figure}

The data in Fig.~2 is obtained by a Gaussian fit to the recorded mw spectra of the MW$_\mathrm{P}$ transition, after various durations $\tau$ of thermal expansion. In Fig.~\ref{pic:SpectroCE}, we plot the recorded spectra for $\tau = 2.8\,\milli\second$ with (blue circles) and without trapping light (red rectangles). The FWHM for untrapped atoms is clearly larger than for trapped atoms. We find $\sigma_{\textrm{P, no trap}} = (0.83 \pm 0.04)\,\mega\hertz$ and $\sigma_{\textrm{P, trap}} = (0.33 \pm 0.02)\,\mega\hertz$. In addition, in the trapped-atom case, we do not observe any pedestal that would indicate a significant untrapped fraction of atoms in the Rydberg cloud.

\medskip
We fit the data in Fig.~2(a) by a simple expansion model that takes into account the gradients of electric field and the thermal motion of the atoms. Considering an initial Gaussian spatial distribution of the Rydberg cloud, we write that, after a free expansion of duration $\tau$,
\begin{align}
	\mean{j^2}(\tau) = \mean{j^2}(\tau=0) + \frac{k_{\mathrm{B}} T}{m_{\mathrm{Rb}}}\times \tau^2,
\end{align}
where $j\in\{x,y,z\}$ and $T$ is the temperature of the atoms. When the Rydberg cloud is freely expanding [red data points in Fig.~2(a)], the expected FWHM of the MW$_\mathrm{P}$ spectrum then reads
\begin{align}
	\sigma_{\textrm{P, no trap}}(\tau) = \sqrt{\sigma_0^2  + 8 \ln{2} (\beta_{\mathrm{Stark}}\boldsymbol{\nabla} F)^2 \frac{k_{\mathrm{B}} T}{m_{\mathrm{Rb}}}\times \tau^2},
\end{align} 
where $(\boldsymbol{\nabla} F)^2 = \left(\partial_x F\right)^2 + \left(\partial_y F\right)^2 + \left(\partial_z F\right)^2$, $\sigma_0$ is the FWHM at time $\tau = 0$, $\beta_\mathrm{Stark} = 99.8\,\mega\hertz\per(\volt\per\centi\meter)$ is the linear Stark shift of $\kell{52}$ and the $8\ln{2}$ factor accounts for the difference between standard deviation and FWHM of a Gaussian profile.  

When we shine the trapping laser, atoms can only move along the $x$-axis. In this case, the FWHM of the MW$_\mathrm{P}$ spectrum reads
\begin{align}
	\sigma_{\textrm{P, trap}}(\tau) = \sqrt{\sigma_0^2  + 8 \ln{2} \left(\beta_{\mathrm{Stark}} \partial_x F\right)^2 \frac{k_{\mathrm{B}} T}{m_{\mathrm{Rb}}}\times \tau^2}.
\end{align}

We fit the data in Fig.2~(a) with the above expressions, $\sigma_0$ being taken as a global fit parameter. We allow for different values of T for each data set and find:
\begin{itemize}
	\item $\sigma_0 = (0.32 \pm 0.01) \mega\hertz$,
	\item $T = (14.6\pm 2.2)\,\micro\kelvin$ for untrapped atoms with $\bar{N} = \bar{N}_0 = 0.08$ (open red rectangles),	
	\item $T = (12.8\pm 0.9)\,\micro\kelvin$ for untrapped atoms with $\bar{N} = 0.18$ (solid red rectangles),
	\item $T = (14.4\pm 3.2)\,\micro\kelvin$ for trapped atoms with $\bar{N} = 0.08$ (open blue circles),
	\item $T = (3.5\pm 1.2)\,\micro\kelvin$ for trapped atoms, with $\bar{N} = 0.18$ (solid blue circles).
\end{itemize}

\section{Numerical simulations of the oscillations in the trap}
We simulate the behavior of the atoms in the trap through numerical Monte-Carlo simulations, in order to reconstruct the observed oscillations in the trap frequency measurements (Fig. 3). 
\begin{figure}[!ht]
\centering
	\includegraphics[width=0.5\textwidth]{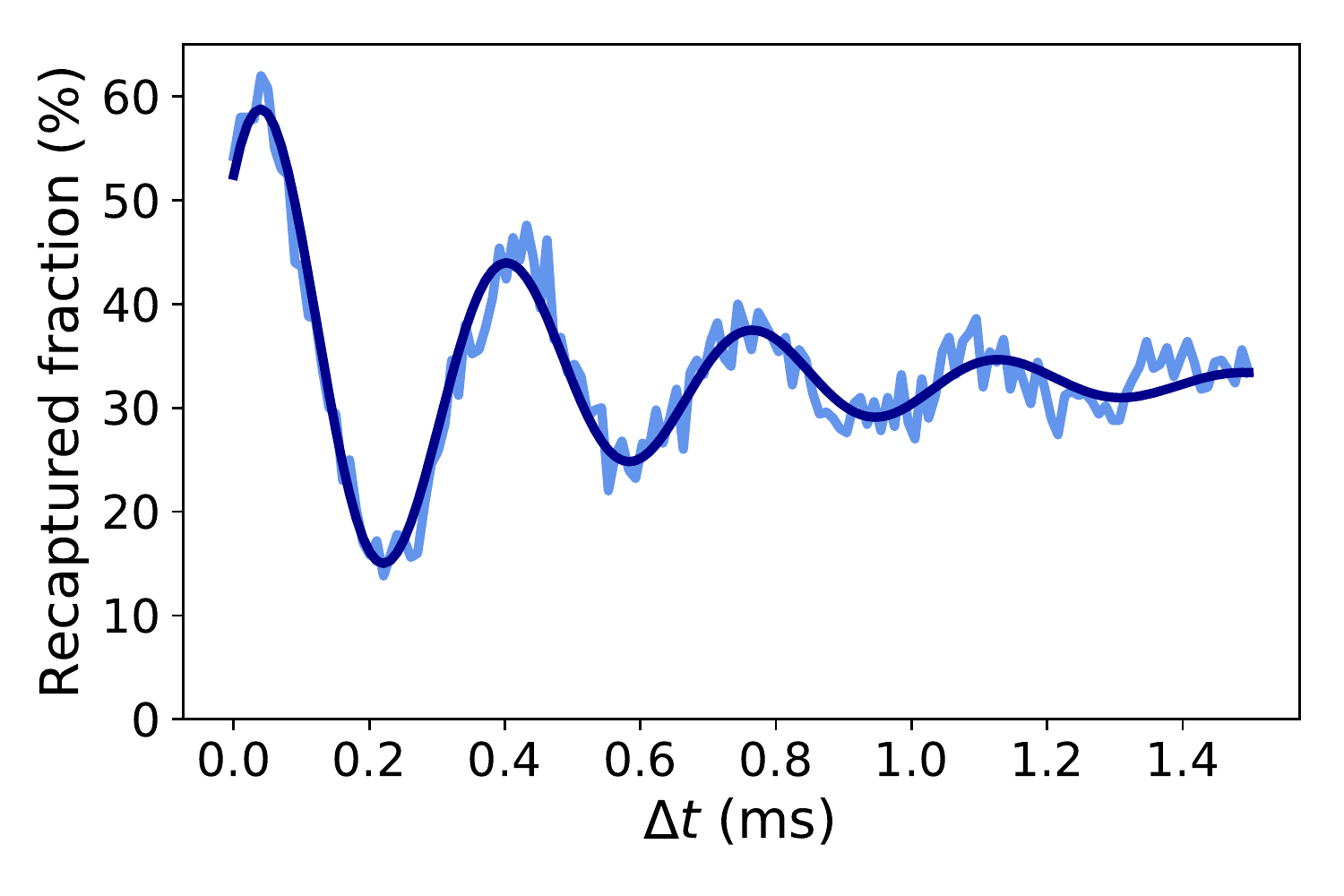}
\caption{Monte-Carlo simulation of the atomic motion in the trap. The solid dark-blue line is a damped sinusoidal fit to the simulated data (light-blue line).}
\label{pic:simu_RR}
\end{figure} 

We calculate the 2D LG-shaped ponderomotive energy potential, using the results of the fit to the data of Fig.~\ref{pic:LS}. We calculate 500 atomic trajectories in the trap, with random initial positions that follow a gaussian distribution, centered on $(y = 0, z= 12\,\micro\meter)$ and with a width of $5\,\micro\meter$. The initial velocities of the atoms are randomly taken according to a Maxwell-Boltzmann distribution of temperature $13\,\micro\kelvin$. We then perform a discrete-time integration of the equations of motion using the odeint routine of Scipy. After the total $T=2.6\,\milli\second$ evolution, we count the fraction of atoms that are still within the trapping region ($\frac{y^2}{w_y^2} + \frac{z^2}{w_z^2} \leq 2$). 

We plot in Fig.~\ref{pic:simu_RR} the result of this simulation. We fit the data by a damped sine and find an oscillation frequency of $(1.14\pm0.01)\,\kilo\hertz$. The difference with the measured frequency of $(1.37\pm0.05)\,\kilo\hertz$ is attributed in part to experimental imperfections. Because of long-term fluctuations of, e.g., the temperature of the SLM or the position of the atom cloud in the cryostat, we need to optimize the shape of the LG beam on a daily basis. We thus expect variations of the intensity distribution from one data-set to the next.

In addition, as mentioned in section~\ref{sec:profile}, the simple model that we use to reconstruct the trapping beam parameters is insensitive to fast variations of the intensity distribution. Such unreconstructed defects of the intensity profile w.r.t.\ the ideal one in Eq.~\eqref{eq:LG_I} can modify the oscillation frequency of the atoms in the trap and thus explain the small difference between the measured and the simulated oscillation frequencies.

\bibliographystyle{plain}

\bibliography{bibliography}